\documentclass[aps,
showpacs,pra,amssymb, amsmath]{revtex4}
\usepackage{amsmath,latexsym,amssymb}
\usepackage[dvips]{graphicx}
\usepackage{amsmath}
\usepackage{latexsym}
\usepackage{amssymb}
\usepackage{bm}

\begin{document}

\title{Realism violates quantum mechanics}

\author{Koji Nagata}
\affiliation{Department of Physics, Korea Advanced Institute of
Science and Technology, Taejon 305-701, Korea}

\date{\today}

\begin{abstract}
Recently, [{arXiv:0810.3134}] is accepted and published.
We show that local realistic theories violate separability of
$N$-qubit quantum states by a factor of $3^N$.
Thus we have a violation factor of $3$ when $N=1$.
Realistic theories 
violate a proposition of a single qubit (a state 
with two-dimensional space)
system, the Bloch sphere machinery of quantum mechanics. 
Our method uses the ratio of each of the scalar products. 
The maximal possible value of the scalar product in local realistic theories grows exponentially larger than one in 
$N$-qubit separable states when the number of qubits increases.
We find the violation factor $3^N$ when the measurement setup is entire range of settings for each of the local observers.
\end{abstract}

\maketitle

Recently, \cite{NagataNakamura} is accepted and published.
Local realistic theory is a set of two 
propositions \cite{bib:Peres3, bib:Redhead}.
One is {\em realism} which states 
that there exist the elements of physical reality
\cite{Einstein}
independent of whether it is observed or not.
Another is Einstein, Podolsky, and Rosen {\em locality} principle which
says that space-like separated measurements
to be mutually independent.
Certain quantum predictions violate Bell inequalities \cite{bib:Bell}, 
which form necessary conditions for local realistic theories for 
the results of measurements.
Thus, some quantum predictions do not allow a local realistic interpretation.

It is well known that all
separable states admit local realistic theories \cite{Werner}.
Recently, Roy derived \cite{Roy} so-called `separability inequalities' and 
Greenberger, Horne, and Zeilinger (GHZ) state \cite{GHZ} violates 
the inequality by a factor of $2^{(N-1)}$.
The assumption to derive the Roy inequality is 
that the system is in separable states.
The inequality can be also derived simply by another quadratic entanglement witness inequality presented in Ref.~\cite{Nagata}.
On the other hand, the GHZ state violates standard Bell 
inequalities \cite{Zukowski,Werner2,Mermin} by a factor of 
$2^{(N-1)/2}$ \cite{Werner2}.
Roy concluded that multiqubit separability inequalities exponentially stronger than standard Bell inequalities by a factor of $2^{(N-1)/2}$.
The Roy work is important since it has revealed the crucial 
difference between the notion of quantum separability and local realistic theories.
That is, local realistic theories violate the Roy inequality
whereas all separable states must satisfy the inequality 
(See also Ref.~\cite{Nagata2}).
However, in the case where $N=1$, a single qubit 
does not violate both standard Bell inequalities and the Roy inequality.

We note here that local realistic theories can only have significance 
for $N$-qubit systems where $N\geq 2$. 
Therefore, we mainly consider realistic theories where $N=1$.
And a single qubit is defined at a single point in space. 
Here, the meaning of
the word `qubit' is `a quantum state with two-dimensional space'.

The current explosive interest in the applications of a single 
photon to linear optical quantum 
information theory \cite{Oliveira,Kim2003,Mohseni2003,NIELSEN_CHUANG,Galindo} prompts us to seek direct qualitative and quantitative signatures of a single qubit system, (a system with two-dimensional state space).
And, realistic theories 
seem to resist implementations of quantum information theory.
Therefore, we consider the relation between a single qubit system and 
realistic theories.

So far, complete understanding between a single qubit system and realistic theories has not been reported. For example, we ask: 
Can realistic theories satisfy all propositions of a single qubit system?
Surprisingly, we shall show that realistic theories 
violate a proposition of a single qubit system, the Bloch sphere.
Thus realistic theories do not satisfy all propositions of 
quantum mechanics.

In this paper, we shall show that 
local realistic theories violate separability of
$N$-qubit quantum states by a factor of $3^N$ 
whereas Roy factor is $2^{(N-1)/2}$.
Thus we have a violation factor of $3$ when $N=1$.
Consequently, realistic theories violate
a proposition of a single qubit system.
One will see that it is the Bloch sphere machinery of quantum mechanics.
Our method uses the ratio of each of the scalar products. 
It turns out that the maximal possible value of the scalar product of correlation functions in local realistic theories grows 
exponentially larger than the value of one in $N$-qubit separable states 
when the number of qubits increases.
The violation factor $3^N$ is obtained in the case where the measurement setup is entire range of settings for each of the local observers.
In contrast, Roy factor $2^{(N-1)/2}$ is obtained in the case where the measurement setup is two orthogonal settings for each of the local observers.

Consider $N$ spin-$1/2$ particles described in pure uncorrelated states.
Each is in a separate laboratory. 
Let us parameterize the local settings
of the $j$th observer with a unit vector $\vec n_j$ with $j=1,\ldots,N$. 
One can introduce the `Bell' correlation function, which is the average of the product of the local results
\begin{equation}
E_{LR}(\vec n_1, \vec n_2,\ldots, \vec n_N) 
= \langle r_1(\vec n_1) r_2(\vec n_2)\ldots r_N(\vec n_N) \rangle_{\rm avg},
\end{equation}
where $r_j(\vec n_j)$ is the local result, $\pm 1$, which is obtained if the measurement direction is set at $\vec n_j$.

Also one can introduce a quantum correlation function with a pure uncorrelated state
\begin{equation}
E_{Sep}(\vec n_1, \vec n_2,\ldots, \vec n_N) = 
{\rm tr}[\rho \vec n_1\cdot \vec \sigma\otimes
\vec n_2\cdot \vec \sigma  \otimes\cdots\otimes  \vec n_N\cdot \vec \sigma]
\label{et}
\end{equation}
where $\otimes$ denotes the tensor product, 
$\cdot$ the scalar product in ${\bf R}^{\rm 3}$, 
$\vec \sigma=(\sigma_x, \sigma_y, \sigma_z)$ is the vector of Pauli operator, 
and $\rho$ is 
a pure uncorrelated state,  
\begin{eqnarray}
\rho=\rho_1\otimes \rho_2\otimes\cdots\otimes\rho_N
\end{eqnarray}
with $\rho_j=|\Psi_j\rangle\langle\Psi_j|$ and $|\Psi_j\rangle$
is a spin-$1/2$ pure state.
One can write the observable (unit) vector $\vec n_j$
in a spherical coordinate system as follows:
\begin{equation}
\vec{n}_j(\theta_j, \phi_j) = \sin \theta_j\cos\phi_j \vec{x}_j^{(1)}
+\sin \theta_j\sin\phi_j \vec{x}_j^{(2)}
+\cos\theta_j \vec{x}_j^{(3)},
\label{vector}
\end{equation}
where $\vec x_j^{(1)}$, $\vec x_j^{(2)}$, and $\vec x_j^{(3)}$
are the Cartesian axes relative to which
spherical angles are measured.

We shall derive a necessary condition to be satisfied by 
the quantum correlation function 
with a pure uncorrelated state given in (\ref{et}).
In more detail, we shall derive the value of the scalar product 
of the quantum correlation function, 
$E_{Sep}$ given in (\ref{et}), i.e., $(E_{Sep}, E_{Sep})$.
We use decomposition (\ref{vector})
and introduce the usual measure
$d\Omega_j=\sin\theta_jd\theta_jd\phi_j$
for the system of the $j$th observer.
We introduce simplified notations as
\begin{eqnarray}
T_{i_1i_2\ldots i_N}=
{\rm tr}[\rho \vec{x}_1^{(i_1)}\cdot \vec \sigma \otimes
\vec{x}_2^{(i_2)}\cdot \vec \sigma  \otimes\cdots\otimes  \vec{x}_N^{(i_N)}\cdot \vec \sigma]
\end{eqnarray}
and
\begin{eqnarray}
\vec c_j=(c^1_{j}, c^2_{j}, c^3_{j})=(\sin \theta_j\cos\phi_j,
\sin \theta_j\sin\phi_j,
\cos\theta_j).
\end{eqnarray}
Then, we have
\begin{eqnarray}
&&(E_{Sep}, E_{Sep})  = 
\int\!\!d\Omega_1\cdots
\int\!\!d\Omega_N\nonumber\\
&&\left(\sum_{i_1,\ldots,i_N=1}^3T_{i_1\ldots i_N}
c^{i_1}_{1}\cdots c^{i_N}_{N}\right)^2 \nonumber\\
&& =  (4\pi/3)^N 
\sum_{i_1,\ldots,i_N=1}^3T_{i_1...i_N}^2\leq (4\pi/3)^N ,
\label{EEvalue}
\end{eqnarray}
where we use the orthogonality relation
$\int d \Omega_j ~ c_j^{\alpha} c_j^{\beta}  
= (4\pi/3) \delta_{\alpha,\beta}$.
From the convex argument, the value of $\sum_{i_1,\ldots,i_N=1}^3T_{i_1\ldots i_N}^2$ is bounded as $\sum_{i_1,\ldots,i_N=1}^3T_{i_1\ldots i_N}^2\leq 1$.
One has
\begin{eqnarray}
\prod_{j=1}^N\sum_{i_j=1}^3 
({\rm tr}[\rho_j \vec{x}_j^{(i_j)}\cdot \vec \sigma])^2\leq 1.\label{Bloch}
\end{eqnarray}
Please notice that the condition (\ref{EEvalue}) is a separability inequality.
All separable states must satisfy the condition (\ref{EEvalue}).
It is worth noting here that
this inequality is only saturated iff $\rho$ is a pure uncorrelated state.
The reason of the derivation of the condition 
(\ref{EEvalue}) is due to the Bloch sphere machinery
\begin{eqnarray}
\sum_{i_j=1}^3 
({\rm tr}[\rho_j \vec{x}_j^{(i_j)}\cdot \vec \sigma])^2\leq 1.
\end{eqnarray}
Thus a violation of the inequality (\ref{EEvalue})
implies a violation of the Bloch sphere machinery.
We have the maximal possible value of the scalar product as 
$(E_{Sep}, E_{Sep})_{\rm max}=(4\pi/3)^N$ when the system is 
in a pure uncorrelated state.

On the other hand, a correlation function satisfies local realistic theories
if it can be written as
\begin{eqnarray}
&&E_{LR}(\vec{n}_1,\vec{n}_2,\ldots,\vec{n}_N)=\nonumber\\
&&\int d\lambda \rho(\lambda)
I^{(1)}(\vec{n}_1,\lambda)I^{(2)}(\vec{n}_2,\lambda)\cdots
I^{(N)}(\vec{n}_N,\lambda),
\label{LHVcofun}
\end{eqnarray}
where $\lambda$ denotes a set of hidden variables, 
$\rho(\lambda)$ is their distribution, and $I^{(j)}(\vec{n}_j,\lambda)$ is 
the predetermined `hidden' result of 
the measurement of all the dichotomic observables 
parameterized by any direction of $\vec n_j$.

We shall derive the maximal possible value of the scalar product 
of the local realistic correlation function, 
$E_{LR}$ given in (\ref{LHVcofun}), i.e., $(E_{LR}, E_{LR})_{\rm max}$.
It will be proven that
\begin{eqnarray}
(E_{LR}, E_{LR})_{\rm max} = (4\pi)^N.
\label{Bell-Zineq}
\end{eqnarray}

A necessary condition for a correlation function in local realistic theories
given in (\ref{LHVcofun})
to satisfy the Bloch sphere machinery
is that the following values of each of scalar products are equal: 
$(E_{LR}, E_{LR})_{\rm max} =(E_{Sep}, E_{Sep})_{\rm max}$.
If one finds $(E_{Sep}, E_{Sep})_{\rm max}< (E_{LR}, E_{LR})_{\rm max}$,
then local realistic theories violate 
the Bloch sphere machinery.
In a single qubit system, realistic theories 
violate the proposition of quantum mechanics.

In what follows, we derive the value of (\ref{Bell-Zineq}).
One has
\begin{widetext}
\begin{eqnarray}
&&\int\!\!d\Omega_1
\cdots
\int\!\!d\Omega_N 
\int d\lambda \rho(\lambda) 
I^{(1)}(\theta_1, \phi_1,\lambda)\cdots
I^{(N)}(\theta_N, \phi_N,\lambda) 
\times \int d\lambda' \rho(\lambda')
I^{(1)}(\theta_1, \phi_1,\lambda')\cdots
I^{(N)}(\theta_N, \phi_N,\lambda')\nonumber\\
&&=\int\!\!d\Omega_1
\cdots
\int\!\!d\Omega_N 
\int d\lambda \rho(\lambda) \int d\lambda' \rho(\lambda')I^{(1)}(\theta_1, \phi_1,\lambda)\cdots
I^{(N)}(\theta_N, \phi_N,\lambda) I^{(1)}(\theta_1, \phi_1,\lambda')\cdots
I^{(N)}(\theta_N, \phi_N,\lambda')\nonumber\\
&&\leq\int\!\!d\Omega_1
\cdots
\int\!\!d\Omega_N 
\int d\lambda \rho(\lambda) \int d\lambda' \rho(\lambda').\label{int}
\end{eqnarray}
\end{widetext}
We have used the following fact: 
\begin{eqnarray}
I^{(j)}(\theta_j, \phi_j,\lambda)
I^{(j)}(\theta_j, \phi_j,\lambda')=\pm 1.
\end{eqnarray}
It is obvious that the inequality (\ref{int}) can be saturated since
\begin{eqnarray}
&&\{\lambda|I^{(j)}(\theta_j, \phi_j,\lambda)=1\}
=\{\lambda'|I^{(j)}(\theta_j, \phi_j,\lambda')=1\},\nonumber\\
&&\{\lambda|I^{(j)}(\theta_j, \phi_j,\lambda)=-1\}
=\{\lambda'|I^{(j)}(\theta_j, \phi_j,\lambda')=-1\}.
\end{eqnarray}
Hence one has
\begin{eqnarray}
&&(E_{LR}, E_{LR})_{\rm max}\nonumber\\
&&=\int\!\!d\Omega_1
\cdots
\int\!\!d\Omega_N
\int d\lambda \rho(\lambda)
\int d\lambda' \rho(\lambda')
 =(4\pi)^N.\nonumber\\
\label{integral}
\end{eqnarray}
Therefore, one has the value (\ref{Bell-Zineq}).
Thus, we have the main result as
\begin{eqnarray}
\frac{(E_{LR}, E_{LR})_{\rm max}}{(E_{Sep}, E_{Sep})_{\rm max}}= 3^N.
\end{eqnarray}
The striking aspect is that realistic theories violate
a proposition of a single qubit system, i.e., the Bloch sphere
machinery when the system is in a pure uncorrelated state.
We have a violation factor $3$ when $N=1$.
Of course, the result does not imply 
that a single-qubit-point cannot admit realistic theories.
That is, if one gets the value of scalar product which equals to some value less than $4\pi/3$,
the results of measurement agree with quantum predictions 
and realistic predictions.
However, our result is still valid because our main 
result is as follows. 
If one takes realism, then the results of measurement do not 
agree with the Bloch sphere machinery---therefore there is 
no such a sphere in space. 
This situation coexists with simulating measurement outcome (of course, the inside of `the Bloch sphere') by realistic theories without the existence of the Bloch sphere.


In conclusion,
we have shown that local realistic theories violate separability of
$N$-qubit quantum states by a factor of $3^N$.
Our analysis has been relied on the ratio of each of the scalar products. 
The maximal possible value of the scalar product in local realistic theories
has increased exponentially larger  than the value of one in 
$N$-qubit separable states when the number of qubits increases.
We have found the violation factor $3^N$ when the measurement setup is entire range of settings for each of the local observers.
Consequently, realistic theories has violated 
a proposition of a single qubit system, i.e., 
the Bloch sphere machinery of quantum mechanics.
However, our result has not implied that a single qubit state
cannot admit realistic theories.

\acknowledgments
This work has been
supported by Frontier Basic Research Programs at KAIST and K.N. is
supported by a BK21 research grant.

\end{document}